\documentstyle[12pt]{article}

\catcode`\@=11

\@addtoreset{equation}{section}
\def\theequation{\arabic{section}.\arabic{equation}}

\def\@normalsize{\@setsize\normalsize{15pt}\xiipt\@xiipt
\abovedisplayskip 14pt plus3pt minus3pt%
\belowdisplayskip \abovedisplayskip
\abovedisplayshortskip  \z@ plus3pt%
\belowdisplayshortskip  7pt plus3.5pt minus0pt}

\def\small{\@setsize\small{13.6pt}\xipt\@xipt
\abovedisplayskip 13pt plus3pt minus3pt%
\belowdisplayskip \abovedisplayskip
\abovedisplayshortskip  \z@ plus3pt%
\belowdisplayshortskip  7pt plus3.5pt minus0pt
\def\@listi{\parsep 4.5pt plus 2pt minus 1pt
            \itemsep \parsep
            \topsep 9pt plus 3pt minus 3pt}}

\def\underline#1{\relax\ifmmode\@@underline#1\else
        $\@@underline{\hbox{#1}}$\relax\fi}
\@twosidetrue

\relax

\catcode`@=12

\catcode`\@=11

\def\section{\@startsection{section}{1}{\z@}{3.5ex plus 1ex minus
   .2ex}{2.3ex plus .2ex}{\large\bf}}

\def\thesection{\Roman{section}.}

\def\appendix{\setcounter{section}{0}
        \def\thesection{Appendix }
        \def\theequation{\Alph{section}.\arabic{equation}}}

\def\ps@headings{\def\@oddfoot{}\def\@evenfoot{}
\def\@oddhead{\hbox{}\hfill
        \makebox[.5\textwidth]{\raggedright\ignorespaces --\thepage{}--
        \hfill {}}}
\def\@oddhead{\hbox{}\hfill --\thepage{}-- \hfill
        {}}
\def\@evenhead{\@oddhead}
\def\subsectionmark##1{\markboth{##1}{}}
}

\ps@headings

\catcode`\@=12

\relax

\def\figcap{\section*{Figure Captions\markboth
        {FIGURECAPTIONS}{FIGURECAPTIONS}}\list
        {Fig. \arabic{enumi}:\hfill}{\settowidth\labelwidth{Fig. 999:}
        \leftmargin\labelwidth
        \advance\leftmargin\labelsep\usecounter{enumi}}}
 \relax
\def\tablecap{\section*{Table Captions\markboth
        {TABLECAPTIONS}{TABLECAPTIONS}}\list
        {Table \arabic{enumi}:\hfill}{\settowidth\labelwidth{Table 999:}
        \leftmargin\labelwidth
        \advance\leftmargin\labelsep\usecounter{enumi}}}
 \relax
\def\reflist{\section*{References\markboth
        {REFLIST}{REFLIST}}\list
        {[\arabic{enumi}]\hfill}{\settowidth\labelwidth{[999]}
        \leftmargin\labelwidth
        \advance\leftmargin\labelsep\usecounter{enumi}}}
 \relax

\catcode`\@=11

\def\ps@headings{\def\@oddfoot{}\def\@evenfoot{}
\def\@oddhead{\hbox{}\hfill
        \makebox[.5\textwidth]{\raggedright\ignorespaces --\thepage{}--
        \hfill {}}}
\def\@evenhead{\@oddhead}
\def\subsectionmark##1{\markboth{##1}{}}
}

\ps@headings

\relax

\newskip\humongous \humongous=0pt plus 1000pt minus 1000pt

\newif\ifdtup

\def\beq{\begin{equation}}
\def\eeq{\end{equation}}

\def\beqn{\begin{eqnarray}}
\def\eeqn{\end{eqnarray}}
\relax

\def\G2{{\; \rm GeV/}c^2}
\def\G{\; \rm GeV}

\def\dotx{\dotx{\dot\overline{x}}}

\relax

\begin{document}
\begin{titlepage}
\begin{flushright}
       {\normalsize    OU-HET 249  \\  hep-th 9609151\\
                   September, 1996 }
\end{flushright}
%
\begin{center} 
  {\large \bf  Infinity Cancellation,  Type $I^{\prime}$  Compactification \\
  and  String ${\cal S}$-Matrix Functional }\footnote{This work is
 supported in part by  Grant-in-Aid for  Scientific Research
(07640403,08640372)
  and by the Grant-in-Aid  for Scientific Research Fund (2126)
from
 the Ministry of Education, Science and Culture, Japan.}

\vfill
         {\bf Y.~Arakane}   \\
         {\bf H.~Itoyama}  \\
         {\bf H.~ Kunitomo} \\
            and \\
              {\bf A.~Tokura}\\
        Department of Physics,\\
        Graduate School of Science, Osaka University,\\
        Toyonaka, Osaka, 560 Japan\\
\end{center}
\vfill
\begin{abstract}

  Nonvanishing  tadpoles and  possible infinities  associated
 in  the multiparticle amplitudes are discussed  with  regard to  the
  disk and $RP^{2}$ diagrams of the Type $I^{\prime}$ compactification.
 We find that the infinity cancellation of $SO(32)$  type $I$ theory 
  extends to this case as well
  despite the presence of tadpoles localized in the D-brane
  world-volume and the orientifold surfaces.
 Formalism of
   string ${\cal S}$-matrix generating functional is presented
 to find a consistent string background as c-number source function:
  we find this only treats the cancellation of the tadpoles in the
 linearized  approximation.
  Our formalism automatically provides  representation of
 the string amplitudes  on this background to all orders in
 $\alpha^{\prime}$.

\end{abstract}
\vfill
\end{titlepage}

\section{Introduction}

There are a few important implications which  the $D$-brane has brought
 to us \cite{Pol2}. In this paper, we  consider one of these which has been
  discussed  relatively little so far.
 The   $D$-brane  can be introduced by taking a $T$-dual of type $I$ theory
 ( or in bosonic string  theory containing both closed and open string
 sectors) \cite{DLP},
 which is  sometimes  referred to as type  $I^{\prime}$.
 The simplest nontrivial example  which will be dealt with here
  is SO(32) Type $I^{\prime}$ theory in flat ten dimensional space
  where  $x^{9}$  is compactified on  a circle with radius $R$.
 This model has 32 Dirichlet  eight-branes at $x^9=0$  forming
  Dirichlet boundary condition of the disk  geometry.
  In addition it has
 two orientifold surfaces at $x^9=0, \pi R$   where  cross cap of the $RP^2$
 geometry  lies.
  The $n$-point  amplitude of the disk diagram and that of the $RP^2$,
  which are next leading to  the spherical geometry in string perturbation
 theory, depend on  these locations respectively.  
These facts can be easily understood by the notion of
 boundary states  in the closed string sector.  ( See Appendix A of this paper
 and \cite{Hor}.)

 It has been observed in ref. \cite{PW}  that  this   model has
 nonvanishing tadpoles localized in spacetime.  This, combined with  dynamics
  captured partly by the low energy  spacetime action \cite{Rom,PW}, exhibits
  breakdown
  of perturbation theory, avoiding an immediate conflict with the heterotic-
  type $I$ duality.

  There are two issues on tadpoles  which are related but which are
 found in this paper to be handled somewhat separately.
   Firstly, the conventional string lore  tells us that nonvanishing
 tadpoles  would in general introduce infinities in
  the multiparticle amplitudes.
  We must, therefore, examine how  the cancellation  of
  $SO(32)$ type $I$  theory  stated in \cite{GS} and established  by
  explicit computation in \cite{IM} may remedy the situation.
  In the next section, the $n$-point amplitude for  the disk and the $RP^{2}$
   diagrams  is computed in the original flat
 background   of  the  type $I^{\prime}$ model defined above.
  We find  that
  the potential infinities appear only in the momentum conserving part
 and  the infinity cancellation  of the Disk and the $RP^{2}$ diagrams
  in type $I$ $SO(32)$  theory  renders the type $I^{\prime}$
  case finite as well.  This is the case
 despite  the presence of the dilaton source  localized at
 the position of the D-brane and that of the orientifold surfaces.

   Secondly,  it is well known that tadpoles are considered to be the source of
 vacuum instability and can be removed by shifting background  geometry.
  This is, of course, the spirit of  the mechanism \footnote{ The original
  mechanism has been discussed in the light-cone string field theory
    and  the $\sigma$  model approach.} illustrated by the
  bosonic string in \cite{FS}.
  We find  that a version of this mechanism   can be implemented
  at the level of string amplitudes   within the first quantized framework.
  In section $3$, we present formalism of string models  based on
  the generating functional  ${\cal S}$ for the $S$-matrix  elements
 \cite{FJKY}.
  This is used to  determine  the background geometry
  as $c$-number source function by  demanding  the cancellation of
 tadpoles.   We also check that this background does not introduce
 new short distance singularity of world sheet through  self-contractions.

  Expansion of ${\cal S}$  around this background automatically produces
   the representation of the string amplitudes to all orders in
 $\alpha^{\prime}$.  This is, however, found to be
 the linearized approximation  to the full nonlinear theory  whose
 low energy behavior  ( leading   in $\alpha^{\prime}$)   is
 captured by   the spacetime action of \cite{Rom,PW}.
  This  and related aspects  are discussed in section $4$. 
   
 In Appendix $A$, we present the disk and  the $RP^{2}$ geometries
 of type $I^{\prime}$  theory as  boundary states.
  In appendix $B$,  we briefly
 discuss  the bosonic example discussed in \cite{FS}  in our formalism.

  It will  be appropriate  to introduce some of our notation here.
 We denote by  $ \left( {\displaystyle \prod_{I=1}^{n} } 
 \zeta_{I}^{M_I N_I} \right)$ 
$\tilde{A}^{(n)}_{{\cal M} \; M_{1} N_{1} \cdots M_{n} N_{n}}( k_{I}^{M_{I}})$
 an $n$-particle amplitude for bosonic massless states of closed string
 with polarization tensor $\zeta_{I}^{M_{I}N_{I}}$ and momentum
 $k_{I}^{M_{I}}$~$I= 1\sim n$.  The contribution from the  worldsheet geometry
   ${\cal M}$ is indicated by  the subscript  ${\cal M}$.
  The well-known formula reads
\beqn
\label{eq:amp} 
 \left( \prod_{I=1}^{n}  \zeta_{I}^{M_I N_I} \right)
\tilde{A}^{(n)}_{{\cal M} \; M_{1} N_{1} \cdots M_{n} N_{n}}
( k_{I}^{M_{I}})~~~~~~~
~~~~~~~~~~ ~~~~~~~~~~~~~~~~~~~~~~~~~~~~~~~~~~~~~~~~~ \nonumber \\
 \equiv  
\int_{(moduli)_{{\cal M}}} d(WP)_{{\cal M}}
 \frac{(det' P_1^{\dagger} P_1)^{1/2}}
{ \tilde{d}  (det' P_{1/2}^{\dagger} P_{1/2})^{1/2}
  vol(Ker P_1)}
\left< \prod_{I=1}^{n} V (k_{I}^{M_{I}} , \zeta_{I}^{M_{I}N_{I}})
 \right>_{{\cal M}}\;\;.
\eeqn
Here the integral in front is the standard integrations over the moduli space
of the Riemann surface ${\cal M}$  and
 $ V (k_{I}^{M_{I}}, \zeta_{I}^{M_{I}N_{I}})$  denotes
  the vertex operator of the particle under question.
We will not ellaborate  upon the notation on this formula here.
 (See \cite{DH}.) $< \cdots >_{{\cal M}}$ means   the functional averaging
 with respect to the action:
\beqn
S[{\bf X}] = \frac{1}{2 \pi \alpha^{\prime}}
 \int_{{\cal M}} d^{2}z d \theta
 d \bar{\theta} \hat{\bf E} \frac{1}{2} 4 \bar{D} {\bf X} D {\bf X} + S_{ct}
 \;\; \nonumber\\
 S_{ct} = \mu^{2} \int_{{\cal M}} d^{2}z \sqrt{\hat{g}}
 + \frac{\ln c \kappa}{4 \pi} \int_{{\cal M}} d^{2}z \sqrt{\hat{g}} R^{(2)}
\eeqn
The second term of $S_{ct}$ of course produces
$(c\kappa)^{-\chi ({\cal M})}$.  Here we take  the loop counting factor
 $c\kappa$ to be proportional to  the string coupling
 $\kappa$\footnote{ The proportionality constant $c$ will be set equal
 to $1$  as this does not spoil the essence of our discussion.}.

    In section $3$, we find it more convenient to  convert
  the expression for the amplitudes into position space
 $A^{(n)}_{{\cal M}\; M_{1} N_{1} \cdots M_{n} N_{n}} ( x_{I})$
   via Fourier transform  albeit being formal:
\beqn
\label{eq:pstn}
 A^{(n)}_{ {\cal M} \; M_{1} N_{1} \cdots M_{n} N_{n}}
 ( x_{1}, \cdots x_{n})
 \equiv  \int \cdots\int   \prod_{I=1}^{n} \frac{d^{D}k_{I}}{(2\pi)^{D}} 
e^{-i k_{I}\cdot x_{I}}  
\tilde{A}^{(n)}_{{\cal M} \; M_{1} N_{1} \cdots M_{n} N_{n}}( k_{I})\;.
\eeqn

The cancellation of the tadpoles against  the  background  is stated as 
\beqn
\label{eq:canc}
\left.
\frac{\delta}{\delta j^{MN}(x)} \left(
 {\cal S}[j^{MN}]  \right) 
\right|_{j^{MN}= \kappa j_{cl}} = 0  \;\;\;.
\eeqn  
  Here $j^{MN}(x)$ is the source function we introduce and
 $\kappa j_{cl}(x)$ is the background geometry we  find.
   The string amplitudes on this background is obtained by simply
 expanding  ${\cal S}[j^{MN}]$  around this point.
   
We collect some of the earlier and the recent references
on orientifolds and scattering off $D$-brane in \cite{sag} and in
 \cite{ori,GMS,scat}.

\section{Type $I^{\prime}$ $Disk$ and $RP^{2}$  Amplitudes 
 in the Original Background and Infinity Cancellation}

 Recall that the  graviton/dilaton vertex  operator  in Type $I$   and
 therefore Type $I^{\prime}$ theory is
\beqn
\label{eq:vo}
 V(k^{M} , \zeta^{MN}) &=&
\kappa \int_{{\cal M}} d^{2}z d \theta
 d \bar{\theta} \zeta^{MN} \int d \bar{\eta}_{N} d \eta_{M}
\nonumber\\
&~&
 : \exp \left( i (k - i \eta D - i \bar{\eta} \bar{D})\cdot {\bf X} +
\frac{\alpha '}{4}   \eta \cdot \bar{\eta}
\bar{\theta} \theta R^{(2)} \right) :
~~~~~~~~~
\eeqn
The second term in the exponent is due to the anomalous contraction of the
dilaton vertex and produces the correct coupling to the two-dimensional
 curvature $R^{(2)}(z, \bar{z})$\cite{FT,deAl}. 
  We  begin with   representing
$\tilde{A}^{(n)}_{disk/RP^{2} \; M_{1} N_{1} \cdots M_{n} N_{n}}
(k_{I}^{M_{I}})$ by the disk and $RP^{2}$ boundary states
 which we denote respectively by $<B \mid$ and
 $<C; R \mid$. The construction of   these states is
 given in Appendix $A$.    Write $\tilde{A}^{(n)}_{disk/RP^{2} \;
 M_{1} N_{1} \cdots M_{n} N_{n}}(k_{I}^{M_{I}} )$ as
\beqn
\prod_{J=1}^{n}  \zeta_{J}^{M_J N_J}
\tilde{A}^{(n)}_{disk/RP^2 \; M_{1} N_{1} \cdots M_{n} N_{n}}( k_{I}^{M_{I}})
& \equiv  &
	\frac{C_{disk/RP^2}'}{ V_{SKV,disk/RP^2}}
	\left<  B/C; R \right|
		 \prod_{I=1}^{n} V(k_{I}^{M} , \zeta_{I}^{MN})
	\left| 0 \right> \;.
\nonumber\\
\;\;
\eeqn
Let us first evaluate the zero mode part. In obvious notation,  we obtain
\beqn 
 {}_{zero}\left<B \right|
	\prod_{I=1}^{n} e^{i k_I \cdot \hat{X}_{zero} }
\left| 0 \right> &\equiv& \prod_{\mu=0}^{8} \otimes 
	{ }_{\mu}\left<B ; p^{\mu}=0 \right|
		e^{i {\displaystyle\sum_{I=1}^{n}} k_I^{\mu} \hat{x}_{\mu} }
	\left| p^{\mu}=0 \right>  \nonumber \\
	& & ~~~~~~~~~~~~~
	\otimes
	{ }_{9}\left<B ;\; x^{9}=x^{9}_B = 0 \right|
		e^{i{\displaystyle \sum_{I=1}^{n}} k_I^{9} \hat{x}_{9} }
	\left| p^{9}=0 \right>
 \nonumber \\
  &=& \left( 
	\prod_{\mu=0}^{8} \delta(\sum_{I=1}^{n} k_I^{\mu}) 
  \right)  \;\;\;,
\eeqn
where $x^{9}_{B}=0$  is the location of  the D-brane world-volume.
   Similarly,
\beqn
& & { }_{zero}\left<C; R \right|
	\prod_{I=1}^{n} e^{i k_I \cdot \hat{X}_{zero} }
\left| 0 \right>
 \nonumber \\
& &= \prod_{\mu=0}^{8} \otimes~_{\mu} 
	\left<C ;\; p^{\mu}=0 \right|
		e^{i  {\displaystyle \sum_{I=1}^{n}} k_I^{\mu} \hat{x}_{\mu} }
	\left| p^{\mu}=0 \right>
 \nonumber \\
	& & ~~~~~~~~~ \otimes	\frac{1}{2}
	\left(
	{ }_{9}\left<C ;\; x^{9}=x^{9}_C =0 \right|
	+ { }_{9}\left<C ;\; x^{9}= \tilde{x}^{9}_C = \pi R \right|
	\right)
		e^{i {\displaystyle \sum_{I=1}^{n}} k_I^{9} \hat{x}_{9} }
	\left| p^{9}=0 \right>
 \nonumber \\
 & &= \left( 
	\prod_{\mu=0}^{8} \delta(\sum_{I=1}^{n} k_I^{\mu}) 
  \right)	\frac{1}{2}
	\left( 1 
	+ e^{i \pi R {\displaystyle \sum_{I=1}^{n}} k_I^{9} }
	\right)\;\;,
\eeqn
where $x^{9}_{C}=0$   and $\tilde{x}^{9}_{C} = \pi R$  are the location
  of  the orientifold surfaces.  The boundary states
 $\left<B \right|$ ~$\left<C; R \right|$ are  eigenstates of the
  total momenta for $M = \mu = 0 \sim 8$ and    those of the
   center of mass  coordinate of string for $M =9$.
  We find
\beqn
\label{eq:antilde}
& & \tilde{A}^{(n)}_{disk/RP^{2} \; M_{1} N_{1}
 \cdots M_{n} N_{n}}(k_{1}^{M} \cdots k_{n}^{M})
  =   \left(
	\prod_{\mu=0}^{8} \delta(\sum_{I=1}^{n} k_I^{\mu}) 
	\right)
      \nonumber\\
 & & ~~~~~~ \times
   \left[  1~ for~B~~or/~~
   \frac{1}{2} \left( 1
		+ e^{ i \pi R {\displaystyle \sum_{I=1}^{n}} k_I^{9} }
 \right) ~for ~C \right]
   \tilde{\tilde{A}}^{(n)}_{disk/RP^{2} \; M_{1} N_{1} 
			\cdots M_{n} N_{n}}(k_{I}^{M})\;\;,  
\nonumber\\
\;\;
\eeqn
  where
\beqn
\tilde{\tilde{A}}^{(n)}_{disk/RP^{2} \; M_{1} N_{1} 
			\cdots M_{n} N_{n}}(k_{I}^{M})
 = \frac{C_{disk/RP^{2}}' \kappa^{n-1}}{V_{SKV,disk/RP^{2}}}
 \prod_{I'=1}^{n}  
   \int d^{2}z_{I'} d \theta_{I'} d \bar{\theta}_{I'}  
   \int d \bar{\eta}_{I'\;N_{I'}} d \eta_{I'\;M_{I'}}  \nonumber \\
\times \exp \left[
                 \pi \alpha' \sum_{I,J}^n  
                     (k_I - i \eta_I D - i \bar{\eta}_I \bar{D})_M
                     (k_J - i \eta_J D - i \bar{\eta}_J \bar{D})_N
 {\bf G}^{MN}_{disk/RP^{2}}(I,J)
  \right. \nonumber \\
 \left.		-\frac{\alpha'}{4} \sum_{I}^n
		 \eta_I  \bar{\eta}_I \theta_{I'} \bar{\theta}_{I'}
		\sqrt{\hat{g}} R^{(2)} \right] \;\;\;.
\eeqn
  where
$C^{\prime} =
 C [\det^{\prime}\Delta]^{-D/2} [\det^{\prime}D]^{D/2}$
  with $ C \equiv [\det^{\prime} P_{1}^{\dagger}P_{1}]^{1/2}
  [\det^{\prime} P_{1/2}^{\dagger}P_{1/2}]^{-1/2}/\tilde{d} $.
   $\tilde{d}$ is the order of the group of diffeomorphism classes and
  $\tilde{d}_{disk} =2$, $\tilde{d}_{RP_{2}} =1$.
 The volume of  the superconformal killing vector  is denoted
  by $V_{SKV,disk/RP^{2}}$.  See \cite{IM,DGW} for more detail.

  It was noticed in ref.~\cite{IM} that, in order to seek for a cancellation
  of infinities in the multiparticle amplitudes
  of $SO(32)$ type $I$ theory, we have to fix the odd
 elements of the invariance group.
 This   is needed as we would like to parametrize the integrand in terms of
 the superspace distance.
 In the operator language, this corresponds to $F_{2}$ picture. 
 The leading divergence is  then of the form
  $ \int \frac{d\lambda}{\lambda} \cdots $  and we can  address the question
 of finiteness via the principal value prescription.
  As in ref. \cite{IM}, we  fix the graded extension of
 $SU(1,1)/ SU(2)$ symmetry of
  the integrand  by setting
 $z_{1}$ and $\arg z_{2}$ and $\theta_{1}$  zero. This offsets the
  volume of the superconformal killing vectors.
  The formula is   then regarded as  the supersymmetric
 extension of the Koba-Nielsen   formula.
  In evaluating this expression, we allow on-shell  condition as well as
  the transversality  to put this in a preferrable form.

  The Green function
 $G^{MN}_{ {\cal M}}(I,J) \equiv G^{MN}_{{\cal M}}
(z_{I}, \bar{z}_{I}, \theta_{I}, \bar{\theta}_{I};  z_{J}, \bar{z}_{J},
 \theta_{J}, \bar{\theta}_{J})$  can be represented  as
\beqn
{\bf G}^{MN}_{disk/RP^{2}}(I,J)
	 &=& {\bf G}^{MN}_{sphere}(I,J) +
 {\bf G}^{MN}_{im,\;disk/RP^{2}}(I,J)
  \nonumber\\
{\bf G}^{MN}_{sphere} &=& 
\eta^{MN} \frac{1}{2 \pi} 
\ln \left| z_I - z_J + i \theta_I \theta_J \right| .
\nonumber\\
{\bf G}^{MN}_{im,\;disk}(I,J)
	&=& (\eta^{\mu \nu} \oplus -\eta^{99})
		\frac{1}{2 \pi} 
		\ln \left|1 - \bar{z}_I z_J \pm \bar{\theta}_I \theta_J \right|
\nonumber\\
{\bf G}^{MN}_{im,\;RP^2}(I,J)
	&=& (\eta^{\mu \nu} \oplus -\eta^{99})
		\frac{1}{2 \pi} 
\ln \left|1 + \bar{z}_I z_J \pm \bar{\theta}_I \theta_J \right| \;\;.
\eeqn
  The sign ambiguity denoted by $\pm$ disappears
 in the final expression.
 It is straightforward to evaluate  the disk and $RP^{2}$  tadpoles
  from the formulas above. In position space, they read
\footnote{ Here we ignore a possible subtlety associated with  the
 compactness of $V_{SKV,RP^{2}}$.}
\beqn 
\label{eq:tadpole} 
 A^{(n=1)}_{ -\chi= 1 \; M N}(x^{M}) &\equiv&
N A^{(n=1)}_{disk \; M N}(x^{M}) + A^{(n=1)}_{RP^2 \; M N}(x^{M})
~, ~~ N=32\;\;,
 \nonumber \\
A^{(n=1)}_{disk \; M N}(x^{M})
 &=& C_{disk}' \frac{r_{disk}}{\pi} \frac{\alpha'}{2}
 \frac{1}{(2\pi)^{9}} \delta(x^9 -x^9_B)
	(\eta_{\mu \nu} \oplus -\eta_{99}) \nonumber \\
A^{(n=1)}_{RP^2 \; M N}(x^{M})
&=& - C_{RP^2}' \frac{r_{RP^2}}{\pi}
	\frac{\alpha'}{2}  \frac{1}{(2\pi)^{9}} 
	\frac{1}{2}
	\left\{
	\delta(x^9 -x^9_C) + \delta(x^9 - \tilde{x}^9_C)
	\right\} \nonumber \\
& &  (\eta_{\mu \nu} \oplus -\eta_{99}) \;\;\;.
\eeqn
  The constants
  $r_{disk}$  and  $r_{RP^2}$    are defined in \cite{IM,DGW}
  and $r_{disk}/r_{RP^2} =2$.    We also quote
\beqn
 C^{\prime}_{disk}/C^{\prime}_{RP^{2}}  = 2^{-(D/2)-1}= 2^{-6} \;\;.
\eeqn
 
 The qualitative feature  of eq.~(\ref{eq:tadpole})  is as is given in
 ref.~\cite{PW}.  Eq.~(\ref{eq:tadpole})     may be associated  with
  the process in which a dilaton/graviton located at $x^{M}$ 
   gets absorbed into  the vacuum.  It is nonvanishing only when
  dilaton/graviton is located in   the  eight-brane  world-volume or 
 the orientifold surfaces.

Let us now turn to the issue of the infinity cancellation.
  First rescale  the worldsheet  Grassmann variables  as
$\theta_I = \lambda^{1/4} \tilde{\theta}_I$~
  in addition to $z_{I} = \sqrt{\lambda} w_{I},~I \geq 2,~ w_{2}=1$,
  so that
$d \theta_I = \lambda^{-1/4} d \tilde{\theta}_I$,
$\frac{\partial}{\partial \theta_I} 
= \lambda^{-1/4} \frac{\partial}{\partial \tilde{\theta}_I}$,
$D_I= \lambda^{-1/4} \tilde{D}_I$.
Next, let
$\eta_I = \lambda^{1/4} \tilde{\eta}_I$
  so that
$\eta_I D_I = \tilde{\eta}_I \tilde{D}_I$
and
$d\eta_I = \lambda^{-1/4} d \tilde{\eta}_I$.
  After these rescalings,   $\tilde{\tilde{A}}_{n}$ is expressible  as
\beqn
\label{eq:anttilde}
\lefteqn{ \tilde{\tilde{A}}^{(n)}_{disk/RP^{2} \; M_{1} N_{1} 
			\cdots M_{n} N_{n}}(k_{1}^{M} \cdots   k_{n}^{M}) }
\nonumber\\
& & = \left( C_{disk/RP^{2}}^{\prime} \right) \kappa^{n-1} r_{disk/RP^{2}}
	\int_0^1 d\lambda  
	\lambda^{-3/2 + \frac{\alpha'}{4}( {\displaystyle\sum_{I=1}^{n} }
 k_I^{9})^2}  
\nonumber\\
& & \times
   \int \prod_{I'=3}^{n} d^{2}w_{I'}
   \int \prod_{J'=2}^{n} d \tilde{\theta}_{J'} d \bar{\tilde{\theta}}_{J'}  
   \int \prod_{K'=1}^{n} d \bar{\tilde{\eta}}_{K'\;N_{K'}} d
 \tilde{\eta}_{K'\;M_{K'}}
 \nonumber\\
& & \times  \exp \left[
                 \pi \alpha' \sum_{I,J}^n  
    (k_I - i \tilde{\eta}_I \tilde{D} - i \bar{\tilde{\eta}}_I 
\bar{\tilde{D}})_M
     (k_J - i \tilde{\eta}_J \tilde{D} - i \bar{\tilde{\eta}}_J
 \bar{\tilde{D}})_N
          \;           {\bf G}^{MN}_{sphere}(I,J) 
                 \right.
 \nonumber\\
& & \;\;\;\;\;\;\;\;\;\;\;\;\;\;\;\;\;
		\left.
		-\frac{\alpha'}{4} \sum_{I}^n
		 \tilde{\eta}_I  \bar{\tilde{\eta}}_I \tilde{\theta}_{I}
 \bar{\tilde{\theta}}_{I}
		\sqrt{g} R^{(2)} \right] 
\nonumber\\
& & \times \left.
	 \exp \left[
                 \pi \alpha' \sum_{I,J}^n  
    (k_I - i \tilde{\eta}_I \tilde{D} - i
 \bar{\tilde{\eta}}_I \bar{\tilde{D}})_M
                     (k_J - i \tilde{\eta}_J \tilde{D} - i
 \bar{\tilde{\eta}}_J \bar{\tilde{D}})_N
          \;           {\bf G}^{MN}_{im,\;disk}(I,J ; \lambda)
                 \right]
     \right|_{\scriptsize \begin{array}{l}w_1=0 \\ w_2=1 \\
 \tilde{\theta}_1=0 \end{array}} \;\;.
 \nonumber\\
\;\;
\eeqn
The factor $ \lambda^{ \frac{\alpha^{\prime} }{4}( {
\displaystyle
 \sum_{I=1}^{n} }
 k_I^{9})^2 }$    comes from the part in  ${\bf G}^{MN}_{sphere}(I,J)$
  which is rescaled to give
\linebreak
 $\eta^{MN} \frac{1}{4\pi} \ln \lambda$.
  As there is no momentum conservation in the ninth direction,
  this
 produces
$\pi \alpha^{\prime}  {\displaystyle \sum_{I,J}^n }
 (k_I - i \tilde{\eta}_I \tilde{D} -
 i \bar{\tilde{\eta}}_I \bar{\tilde{D}})_M  (k_J - i \tilde{\eta}_J 
\tilde{D} - i \bar{\tilde{\eta}}_J \bar{\tilde{D}})^M$
 $\frac{1}{4\pi} \ln \lambda$
 $= \frac{\alpha^{\prime} }{4}( {\displaystyle \sum_{I=1}^{n} }
 k_I^{9})^2 \ln \lambda$.

    Let us discuss  the Grassmann integrations  over
   $\tilde{\theta}$,~ $\tilde{\eta}$ (analytic  variables)
  and $\bar{\tilde{\theta}}$~$\bar{\tilde{\theta}}$ (anti-analytic  variables).
 The leading divergence comes from the case in which one picks up as many as 
 possible  terms from  the first exponent containing
    ${\bf G}^{MN}_{sphere}(I,J)$. After doing this, one is left with
one analytic variable and one antianalytic variable, which  need to be
 saturated  by  a term   from the second exponent.
  This term contains $\lambda^{1/2}$.
   We can  therefore write eq.~(\ref{eq:anttilde}) as
\beqn
\lefteqn{ \tilde{\tilde{A}}^{(n)}_{disk \; M_{1} N_{1} 
			\cdots M_{n} N_{n}}(k_{1}^{M} \cdots   k_{n}^{M}) }
\nonumber\\
& & =   
 C_{disk}^{\prime} \kappa^{n-1} r_{disk}
	\Biggl(
	\int_0^{1} d\lambda 
	\lambda^{-1 + \frac{\alpha'}{4}({\displaystyle\sum_{I=1}^{n} }
 k_I^{9})^2}
  F_{ M_{1} N_{1}\cdots M_{n} N_{n}}^{(n)}
( \lambda ; k_{1}^{M_{1}} \cdots   k_{n}^{M_{n}})
\nonumber\\
& & ~~~~~~~~~~~~~~~~~~~~~~~~~~~~~~ +
   ~{\rm higher~order~terms~in}~ \lambda \Biggr)\;\;\;.
\nonumber\\
\;\;
\eeqn
  Here  $F_{ M_{1} N_{1}\cdots M_{n} N_{n}}^{(n)}$ is some function
   regular and nonvanishing at $\lambda =0$.
   Similarly, we find for $RP^{2}$
\beqn
\lefteqn{ \tilde{\tilde{A}}^{(n)}_{RP^{2} \; M_{1} N_{1}
                        \cdots M_{n} N_{n}}(k_{1}^{M} \cdots   k_{n}^{M})}
\nonumber\\
& & =
 C_{RP^{2}}' \kappa^{n-1} r_{RP^{2}}
         \Biggl(
        \int_0^{1 } d\lambda
        \lambda^{-1 + \frac{\alpha'}{4}({\displaystyle\sum_{I=1}^{n} }
 k_I^{9})^2}
  F_{ M_{1} N_{1}\cdots M_{n} N_{n}}^{(n)}
(- \lambda ; k_{1}^{M_{1}} \cdots   k_{n}^{M_{n}})
\nonumber\\
& & ~~~~~~~~~~~~~~~~~~~~~~~~~~~~~~ + ~{\rm higher~order~terms~in}~ \lambda
 \Biggr) \;\;\;.
\eeqn
  We find that the individual amplitude
 $\left( {\displaystyle \prod_{I=1}^{n} }  \zeta_{I}^{M_I N_I} \right)
\tilde{A}^{(n)}_{disk/RP^{2} \;
 M_{1} N_{1} \cdots M_{n} N_{n}}( k_{I}^{M_{I}})$
 is infinite only when   the sum
  ${\displaystyle  \sum_{I =1}^{n} k_{I}^{9} }$ vanishes.
  But in this region, the dependence  on  $x_{C~9}= \pi R$
   coming from the zero mode integrations ( see eq.~(\ref{eq:antilde}) )
  becomes irrelevant.  The original infinity cancellation of
 $SO(32)$ type $I$  theory persists under type $I^{\prime}$ compactification. 
 We can, therefore, write as
\beqn
   N \lefteqn{
 \left( {\displaystyle \prod_{I=1}^{n} }  \zeta_{I}^{M_I N_I} \right)
\tilde{A}^{(n)}_{disk \;
 M_{1} N_{1} \cdots M_{n} N_{n}}( k_{I}^{M_{I}})
 +
\left( {\displaystyle \prod_{I=1}^{n} }  \zeta_{I}^{M_I N_I} \right)
\tilde{A}^{(n)}_{ RP^{2} \;
 M_{1} N_{1} \cdots M_{n} N_{n}}( k_{I}^{M_{I}})
}
\nonumber \\
& &  =  C_{disk}^{\prime} \kappa^{n-1} r_{disk}
  \left[ N \int_{0}^{1} \frac{d \lambda}{\lambda} F_{n}(\lambda)
  -32 \int_{0}^{1} \frac{d \lambda}{\lambda} F_{n}(-\lambda)
   + ~{\rm finite~terms}~ \right]\;\;\;,
\eeqn
  where $N=32$, $F_{n}(\lambda)$ is some function nonvanishing
 and regular at $\lambda =0$ and the factor $32$ is accounted for by  
 $(r_{disk}/r_{RP^2}) \cdot (C^{\prime}_{disk}/C^{\prime}_{RP^{2}})= 2^{-5}$.
  Tadpoles are nonvanishing locally but, for infinity cancellation,
 it is sufficient for them to cancel in the whole space.

\section{ String ${\cal S}$-Matrix Functional and Consistent Background}
 
  We have seen  that theory is finite despite  the presence of
  the localized tadpole sources.  The  presence of
 nonvanishing tadpoles  itself, however, implies  instabilily of the
 vacuum we have chosen to work with, namely, the geometry of flat spacetime.
   This is supported by the spacetime action  of \cite{Rom,PW}

A nontrivial background must be found which is consistent with  the
 propagation of strings and which offsets  tadpoles in the original background.
 This shift of background can be discussed  within the first quantized
 framework.
 
Let us first introduce  the ${\cal S}$ matrix generating functional:
\beq
\label{eq:calS}
{\cal S} [j^{MN}] \equiv
 \sum_{topologies} {\cal S}_{{\cal M}} [j^{MN}]\;\;\;,
\eeq
  where
\beqn
\label{eq:calSM}
 {\cal S}_{{\cal M}} [j^{MN}]
      \equiv \sum_{n=0}^{\infty} \frac{1}{n!}
            \int \cdots \int \prod_{I=1}^{n} d^{D}x_{I}
         \left( \prod_{J=1}^{n} j^{M_{J} N_{J}}(x_{J}) \right)
A^{(n)}_{{\cal M}\; M_{1} N_{1} \cdots M_{n} N_{n}}
 ( x_{1} \cdots x_{n})\;\;\;.
\eeqn
The indices $(M,N,\cdots)$ run $0 \sim D-1$~with~$D=10$.  Here  $j^{MN}(x)$  is
 a source function  which we introduce in place of  a polarization
 tensor, which is an external wave function. Its Fourier transform is denoted
 by
\beqn
 \tilde{j}^{MN}(k) \equiv \int d^{D}x e^{-ik\cdot x} j^{MN}(x) \;\;\;.
\eeqn

The spherical topology has vanishing  $n$-point amplitudes for $n \leq 2$.
For our purpose, we  find it necessary to
  add  by hand  to the original expression the two point  amplitude
\beqn
\label{eq:2pt}
\tilde{A}^{(2)}_{ sphere \; M_{1} N_{1} M_{2} N_{2}}(k_{I})
=  \eta_{M_{1} N_{1}} \eta_{M_{2} N_{2}} k_{I}^{2}\;\;,
\eeqn
  which is of course the inverse propagator  and vanishes on-shell.
  This is, however, an important ingredient to our discussion.
 
   From eqs.~(\ref{eq:calSM}) and (\ref{eq:vo}), we find
\beqn
\label{eq:calSformula}
 {\cal S}_{{\cal M}}[j^{MN}] &=&
\int_{(moduli)_{{\cal M}}} d(WP)_{{\cal M}}
 \frac{(\det' P_1^{\dagger} P_1)^{1/2}}
{  \tilde{d} (\det' P_{1/2}^{\dagger} P_{1/2})^{1/2}
  vol(\ker P_1)}   \nonumber\\
 &~& \times \left< \exp \left[ \kappa \int_{{\cal M}} d^{2}z d \theta
 d \bar{\theta}
   \hat{\bf E} j_{MN} \left( {\bf X}(z, \bar{z},
  \theta, \bar{\theta}) \right) 
 \bar{D} {\bf X}^{N} D {\bf X}^{M} \right. \right. \nonumber \\
 & & ~~~~~~~~~~~~~ + \left. \left.
 \frac{ \alpha^{\prime} \kappa }{4} \int_{{\cal M}} d^{2}z 
 \sqrt{\hat{g}} j^{M}_{M} \left( X(z, \bar{z}) \right)   R^{(2)}
 \right] \right>_{{\cal M}}\;~~~~ , 
\eeqn 
  where
\beq
 j^{MN} \left({\bf X} \right) \equiv
 \int \frac{d^{D}k }{(2\pi)^{D}}    \exp \left( ik \cdot {\bf X} \right)
 \int d^{D}x  e^{-ik \cdot x} j^{MN}(x)\;\;.
\eeq
  See \cite{DH}  for the rest of  the notation in eq.~(\ref{eq:calSformula}). 
  The salient feature of eq.~(\ref{eq:calSformula})
 is that  the dynamical variable
   ${\bf X}$ ends up with appearing in the argument of $j^{MN}$.
   That ${\bf X}$ appears only  through the exponential operator is
   in accordance with the paradigm of conformal field theory.  
 The second term in the  exponent  originates from the curvature term
  in the dilaton vertex operator of eq.~(\ref{eq:vo}). This identifies
 $j^{M}_{M} \left( X(z, \bar{z}) \right)$  with dilaton field. 
 On-shellness can be implemented  by
 inserting the on-shell delta function $\delta(k^{2})$ in
 $j^{MN}\left({\bf X}\right)$:
\beq
 j^{MN}_{(mod)} \left({\bf X} \right) \equiv
 \int \frac{d^{D}k }{(2\pi)^{D}}  
 \delta(k^{2})  \exp \left( ik \cdot {\bf X} \right)
 \int d^{D}x  e^{-ik \cdot x} j^{MN}(x)\;\;.
\eeq
  Expansion around nonzero value of $j^{MN}$ is regarded as the one around
 the nontrivial string background: the expression inside $< \cdots >$
 agrees with the standard $\sigma$ model expression. 
 
 The cancellation of  the tadpole amplitude 
of the $SO(32)$  type $I$ superstring  up to $ -\chi(M)\leq -1 $ is stated as
\beqn
\left.
\frac{\delta}{\delta j^{MN}(x)} \left(
 {\cal S}_{sphere}[j^{MN}]  + {\cal S}_{disk}[j^{MN}]
  + {\cal S}_{RP^{2}}[j^{MN}] \right) 
\right|_{j^{MN}=0} = 0  \;\;\;.
\eeqn  
 This implies that the flat ten-dimensional Minkowski space is perturbatively
  stable  for $SO(32)$.

  As stated in introduction,  we are concerned with 
 equation\footnote{We are assuming power counting by $\kappa$, so that
  the consideration here is genus by genus  argument.}
\beqn
\left.
    \frac{\delta}{\delta j^{MN}(x)}
\left( {\cal S}_{sphere}[j^{MN}] + {\cal S}_{disk}[j^{MN}] +
 {\cal S}_{RP^{2}}[j^{MN}] \right)
 \right|_{j^{MN}(x) = \kappa j_{cl}^{MN}(x)} = 0 \;\;.
\label{eq:ct}
\eeqn
In position space, eq.~(\ref{eq:ct}) is
\beq
\frac{\partial}{\partial x_{L}}
 \frac{\partial}{\partial x^{L}} j_{cl}^{MN}(x^{M}) =  A_{disk}^{MN}(x^{M}) +
 A_{RP^2}^{MN}(x^{M})\;\;.
\label{eqn:cancel_tad_pos}
\eeq
In momentum space, it reads  as
\beq
 k^2 \tilde{j}^{MN}_{cl}(k) +\tilde{A}_{disk}^{MN}(k) +
\tilde{A}_{RP^2}^{MN}(k) = 0\;\;.
\label{eq:ctadm}
\eeq
  Note that
  we are extending, albeit minimally, the expression
 off-shell.
  In order to check the validity of this recipe,
  we consider in Appendix $B$ the example \cite{FS} of  the bosonic string,
where   we reproduce  the argument of \cite{FS}  for the cancellation of
  the torus tadpole by  the background on spherical topology. 

 The solution to eq.~(\ref{eqn:cancel_tad_pos}) is 
\beqn
\label{eq:sol}
j_{cl}^{MN}(x^9) = \alpha^{\prime} a_{1}(\eta_{\mu \nu} \oplus -\eta_{99})
 \{ x^9 \; \Theta(x^9) -(x^9 - \pi R)\; \Theta(x^9 - \pi R)\}  \;\;\;,
\eeqn 
  where $ a_{1} =  C_{disk}^{\prime} \frac{r_{disk}}{\pi}8
\frac{1}{(2\pi)^{9}}$ and 
$\Theta(x^9)$ is  the step function.
  In momentum space,
\beq
\tilde{j}_{cl}^{M N} (k^{9})
   = - \alpha^{\prime} a_{1}
(\eta_{\mu \nu} \oplus -\eta_{99}) \frac{1}{(k^{9})^2}
	 \left[ 1 - e^{+i \pi  k^{9} R} \right]
	\left( \prod_{\mu=0}^{8} \delta( k^{\mu}) \right) \;\;\;.
\eeq

  Let us finally check  that  this background determined does not create
 any new short distance singularity off-shell which may arise
  from self-contractions of the operators.
  This is necessary to keep consistency with the conclusion of section $2$.  
 The $n$-point amplitude on this  background  is expressed as
\beqn
  F.T.~
\left.
\frac{\delta^{n}}{\delta j^{M_{1}N_{1}}(x_{1}) \cdots 
 \delta j^{M_{n}N_{n}}(x_{n}) } \left(
 {\cal S}[j^{MN}]  \right) 
\right|_{j^{MN}= \kappa j_{cl}}  \;\;\;,
\eeqn  
  where $F.T.$ denotes  Fourier transforms  over $x_{1}, \cdots x_{n}$.
  This contains  the original sphere $n$-point amplitude of the flat spacetime
  at $\kappa^{n-2}$ .  At $\kappa^{n-1}$,  it contains the disk and the
 $RP^{2}$ $n$-point of the flat spacetime and the sphere $(n+1)$-point  where
 one of the spacetime arguments  is integrated together with this background.
 We are concerned with  this latter quantity  which reads
\beqn   
&F.T.& \int d^{10} x_{n+1} A^{(n+1)}_{sphere \; M_{1} N_{1} 
\cdots M_{n+1} N_{n+1}}(x_{1}^{M}, \cdots ,x_{n+1}^{M})
	 j_{cl}^{M_{n+1} N_{n+1}} (x_{n+1}^{M})
  \nonumber \\ 
& &  = ~
\int  \left( d^{9}k_{n+1}  \sum_{k_{n+1}^{9}}
	\right)
	\tilde{j}_{cl}^{M_{n+1} N_{n+1}} (k_{n+1}^{9})
	\left(
	\prod_{M=0}^{9} \delta(\sum_{I=1}^{n+1} k_I^{M}) 
	\right)
\nonumber\\
& & ~~~~~~~~~~~~~~~~~~~~~~~~~~~~~~~~~~~~~~~~~~ \times
	\tilde{\tilde{A}}^{(n+1)}_{sphere \; M_{1} N_{1} 
		\cdots M_{n+1} N_{n+1}}(k_{1}^{M}, \cdots ,k_{n+1}^{M})
  \;\;.  ~~~~~
\eeqn
 Carrying out the momentum integrations and the sum for $k_{n+1}^{M}$, we find
\beqn
\label{eq:3.17}
& & 
  = - \alpha^{\prime} a_{1}
(\eta^{\mu_{n+1} \nu_{n+1} } \oplus -\eta^{99}) \frac{1}
{ \left({\displaystyle \sum_{I=1}^{n} } k_{I}^{9} \right)^2 }
	 \left[ 1 - e^{-i
 \pi \left(  {\displaystyle \sum_{I=1}^{n} } k_{I}^{9} \right) R } \right]
\nonumber \\
& & \times \left(
	 \prod_{\mu =0}^{8} \delta(\sum_{I=1}^{n} k_I^{\mu}) 
	\right)
	\tilde{\tilde{A}}^{(n+1)}_{sphere \; M_{1} N_{1} 
		\cdots M_{n+1} N_{n+1}}(k_{1}^{M}, \cdots ,
k_{n+1}^{M} =  - \delta^{M,9} \sum_{I=1}^{n}k_{I}^{9}  )
 \;\;\;.
\eeqn
    There are now potential  infinities coming from
  short distance singularity of the worldsheet. This  is because
the $k_{n+1}^{M}$ is not put on shell.  The divergent part of
 $\tilde{\tilde{A}}^{(n+1)}_{sphere \; M_{1} N_{1} \cdots M_{n+1} N_{n+1}}$
 multiplied by 
 $1/({\displaystyle \sum_{I=1}^{n} } k_{I}^{9})^2 $
 is proportional to
\beqn
& &\frac{1}{ ({\displaystyle \sum_{I=1}^{n} } k_{I}^{9})^2 }
  \exp \left( \pi \alpha^{\prime} \sum_{I=1}^{n+1} k_{I} \cdot k_{I}
 \frac{1}{2\pi} \ln \epsilon \right) \nonumber \\
 &=&
 \frac{1}{ ({\displaystyle \sum_{I=1}^{n} } k_{I}^{9})^2 }
 \epsilon^{ \frac{\alpha^{\prime}}{2}
 \left( {\displaystyle \sum_{I=1}^{n} } k_{I}^{9}\right)^{2} }
= \frac{\alpha^{\prime}}{4} \int^{\epsilon^{2}}_{0} d \lambda
\lambda^{ -1 + \frac{\alpha^{\prime}}{4}
 \left( {\displaystyle \sum_{I=1}^{n} } k_{I}^{9}\right)^{2}  } \;\;\;.
\eeqn 
  Here $\epsilon$ is an invariant short distance cutoff.
 This  expression is  the same as  the one seen in the last section
 and is singular as $\epsilon \rightarrow 0$  only when
\beq
 \sum_{I=1}^{n} k_{I}^{9} =0\;\;. \nonumber
\eeq
 This regime, however, is killed by the prefactor
$ \left( 1 - e^{-i\pi \left(  {
\displaystyle
 \sum_{I=1}^{n} } k_{I}^{9} \right) R} \right) $  seen in eq.~(\ref{eq:3.17})
 and  this is what we wanted to show.

\section{Discussion} 

 The cancellation of infinities and  the removal of tadpoles are
 two basic constraints for consistent type $I$  string vacua.  In  the case
  of type $I^{\prime}$, we find that these two  are established separately
 without further condition. 
  This tells us naturalness of introducing $D$-branes.

  On the other hand,  the nontrivial background  we have just found in
 eq.~(\ref{eq:sol}) must be compared with the solution \cite{PW}
  from  the low-energy spacetime action.
  This latter one reads
\beqn
\label{eq:PWsol}
\frac{\partial}{\partial x^{9}}  \frac{\partial}{\partial x^{9}} 
  Z(x^{9}) &=& - 24 \sqrt{2} C \mu_{8} \delta(x^{9}) \;\;\;, \nonumber \\
 e^{\phi(x^{9})} &=&  Z(x^{9})^{-5/6}\;, \; \Omega(x^{9})= C 
Z(x^{9})^{-1/6}\;, \; g_{MN} = \Omega \eta_{MN}\;\;, \\
\mu_{8} &=& (2\pi)^{-9/2} (\alpha^{\prime})^{-5/2} \;\;\;. \nonumber
\eeqn
  We see  that our solution from the string amplitudes
 is only a linearized approximation to eq.~(\ref{eq:PWsol}).
  To regain eq.~(\ref{eq:PWsol}), we will have to  add infinite number of
 terms to ${\cal S}$  which vanish on-shell $k^{2}=0$.
  This is of course consistent with the derivative self-interaction of
  the dilaton in the spacetime action \footnote{This is in contrast  to 
 the case of tachyon where
  the  tree level  string ${\cal S}$ matrix functional 
  in the zero-slope limit obeys the same non-linear equation as  that of
 $\phi^{3}$  field  theory. (See the third ref. of \cite{FJKY}.) }.
 
The construction of string theory
  as is currently formulated  provides a set of prescriptions to
 compute  on-shell scattering amplitudes.
 The local Weyl invariance, which is a guiding principle of
 critical string theory, uses explicitly on-shellness.
  The point raised by eqs.~(\ref{eq:sol}) and (\ref{eq:PWsol}), therefore,
  confronts ourselves to a limitation  of our recipe to the  string dynamics
  which we   would like to formulate.
  The hope is  that  the issues discussed in this paper  will,
 at the same time, navigate us somewhere, teaching what string theory ought
 to be.
           

\section{Acknowledgements}
  We thank Nobuyuki Ishibashi and Joe Polchinski  for useful discussion
  on this subject.  Two of us (H.K. and A.T.) acknowledge Kashikojima
  Institute for hospitality.


\newpage
\appendix
\section{A}
The presence of $D$-brane  as well as  that of orientifold  surfaces
  can be captured  by   the notion of boundary  state.
In type $I^{\prime}$ theory  with  circle of radius $R$ in the ninth
  direction   which is  considered in the text,
 the disk and $RP^{2}$  diagrams  are  respectively associated with
  the boundary state $<B|$  and   the boundary  crosscap state $<C|$ 
in the closed string sector. 
  We only need the bosonic sector  here.
They are defined by
\beqn
& &<B| \left( X^9(z, \bar{z})  - 2 \pi n R  \right) \mid_{\tau
= 0} = 0 \;\;,\;\; n \in {\cal Z} \;\;, \;\; -\pi < \sigma < \pi
 \;\;\;, \nonumber \\
& & <B| \frac{\partial}{\partial \tau} X^{\mu}
 (z, \bar{z})\mid_{\tau= 0} = 0 \;\;,
\label{con;boundary}
\\
& &<C| \left(  X^9(z, \bar{z})+  X^9(-1/z, -1/\bar{z})
 - 2 \pi n R \right) \mid_{\tau
= 0} = 0 \;\;,\;\; n \in {\cal Z} \;\;, \;\; 0 < \sigma < \pi \;\;\;,
 \nonumber \\
& &<C| \left(  X^{\mu} (z, \bar{z})
 -  X^{\mu} ( -1/z, -1/\bar{z})  \right) \mid_{\tau= 0} = 0 \;\;.
\label{con;crosscap}
\eeqn
where $ z= e^{\tau + i\sigma}$.
The mode expansion of  closed string coordinate is
  as usual
\beq
X^M (z, \bar{z}) = X_R^M (z) + X_L^M (\bar{z}) \;\;\;\; \;\;,
\eeq
where
\beqn
X_R^M (z) &=& \frac{1}{2} x^M
                - i \frac{\alpha'}{2} p_R^M \ln z
                + i \sqrt{\frac{\alpha'}{2}}
          \sum_{n \neq 0} \alpha_n^M \frac{z^{-n}}{n} \;\;,\nonumber \\
X_L^M (\bar{z}) &=& \frac{1}{2} x^M 
                - i \frac{\alpha'}{2} p_L^M \ln \bar{z}
                + i \sqrt{\frac{\alpha'}{2}}
       \sum_{n \neq 0} \tilde{\alpha}_n^M \frac{\bar{z}^{-n}}{n}\;\;, 
\eeqn
and  $p^M =( p_R^M + p_L^M )/2 $,  $ \ell^M =( p_R^M - p_L^M )/2 $.
 
In terms of modes,  eqs.~(\ref{con;boundary}) and (\ref{con;crosscap}) read
\beqn 
& &<B|\ell^{9}=0 \;,\;<B|(x^{9} - 2 \pi n R )=0\;,\;
<B| ( \alpha_n^{9} - \tilde{\alpha}_{-n}^{9} )=0\;, \nonumber \\
& &<B|p^{\mu}=0 \;,\;
<B| ( \alpha_n^{\mu} + \tilde{\alpha}_{-n}^{\mu} )=0\;,\;
\label{con;mode_boundary}
 \\
& &<C|\ell^{9}=0 \;,\;<C|(x^{9} - \pi n R )=0\;,
<C| ( \alpha_n^{9} - (-1)^n \tilde{\alpha}_{-n}^{9} )=0\;,\; \nonumber \\
& &<C|p^{\mu}=0 \;,\;
<C| ( \alpha_n^{\mu} + (-1)^n \tilde{\alpha}_{-n}^{\mu} )=0 \;,\;
\label{con;mode_crosscap}
\eeqn
 Note that, in contrast with type $I$ case, both $<B|$ and $<C|$
  are eigenstates of the center of mass coordinate $x^{9}$ as opposed
 to   the total momentum $p^{9}$.  Their eigenvalues represent
  the position of $D$-brane and  the orientifold surfaces respectively.
  Eqs.(\ref{con;mode_boundary}) and (\ref{con;mode_crosscap}) are solved
 to give 
\beqn
\left<B \right|
 &=& { }_{zero}\left<B \right|
\otimes { }_{osci}\left< 0 \right|
\exp \left[- \sum_{m>0} \frac{1}{m} \alpha_{\mu \; m} \tilde{\alpha}_{m}^{\mu}
	  + \sum_{m>0} \frac{1}{m} \alpha_{9 \; m} \tilde{\alpha}_{m}^{9}
	\right] \;\;,
\nonumber\\
\left<C \right|
 &=& { }_{zero}\left<C \right|
\otimes { }_{osci}\left< 0 \right|
\exp \left[- \sum_{m>0} \frac{(-1)^m}{m} \alpha_{\mu \; m}
 \tilde{\alpha}_{m}^{\mu}
	  + \sum_{m>0} \frac{(-1)^m}{m} \alpha_{9 \; m} \tilde{\alpha}_{m}^{9}
	\right]\;\;,
\eeqn
\beq
{ }_{zero}\left<B /C \right|
\equiv \prod_{\mu=0}^{8} \otimes 
	{ }_{\mu}\left<B /C; p^{\mu}=0 \right|
\otimes
	{ }_{9}\left<B /C;\; x^{9}=x^{9}_{B / C}  \right| \;\;,
\eeq
  with $x_{B}^{9} =0$ and $x_{c}^{9} = 0, \pi R$.

The Green's functions used in the text  are obtained by evaluating
\beq
{\bf G}^{MN}_{disk/RP^{2}}(z_{I}, \bar{z}_{I}, \theta_{I}, \bar{\theta}_{I};
  z_{J}, \bar{z}_{J}, \theta_{J}, \bar{\theta}_{J})
 = D_I \bar{D}_I [ \tilde{G}^{ MN}_{disk/RP^{2}}
(z_{I}, \bar{z}_{I};  z_{J}, \bar{z}_{J}) \; 
(\theta_{I} - \theta_{J}) (\bar{\theta}_{I} - \bar{\theta}_{J}) ]
\eeq
where the bosonic part $\tilde{G}^{ MN}_{disk/RP^{2}}$ is given by
\beqn
 (-2 \pi \alpha^{\prime})
 \tilde{G}^{ MN}_{disk/RP^{2}} = <B/C| X^M(z_{I}, \bar{z}_{I})
 \; X^N(z_{J}, \bar{z}_{J}) |0>  - ~_{zero}\left<B /C \right|
 x^M x^N \left| 0 \right>
\eeqn

\section{B}

The $S$-matrix functional  of the closed bosonic string
 for the dilaton at tree and one-loop level  reads  as
\beqn
 {\cal S}_{sphere} [j]
&= &
- \frac{1}{2} \int d^{D}x j(x) \frac{\partial}{\partial x^{M}} 
\frac{\partial}{\partial x_{M}}
 j(x)
\nonumber\\
& & + \sum_{n=0}^{\infty} \frac{1}{n!}
            \int \cdots \int \prod_{I=1}^{n} d^{D}x_{I}
         \prod_{J=1}^{n} j(x_{J})
          A^{(n)}_{sphere} ( x_{1} \cdots x_{n} )
 \\
 {\cal S}_{ torus} [j]
&=& \sum_{n=0}^{\infty} \frac{1}{n!}
            \int \cdots \int \prod_{I=1}^{n} d^{D}x_{I}
         \prod_{J=1}^{n} j(x_{J})
          A^{(n)}_{torus } (  x_{1} \cdots x_{n})
\eeqn
  Eq.~(\ref{eq:canc}) is
\beq
\left.  
  \frac{\delta}{\delta j(x)} 
	\left( {\cal S}_{sphere} [j] + {\cal S}_{torus} [j] \right)
 \right|_{ j = \kappa_{dil} j_{cl}(x)} = 0 ,
\eeq
  which determines $j_{cl}(x)$:
\beq
- \kappa_{dil}
\frac{\partial}{\partial x^{M}} \frac{\partial}{\partial x_{M}}
 j_{cl}(x) + A_{torus}^{(1)}(x)  = 0 
\label{eqn:fs_tadpole_cancel}
\eeq
The dilaton vertex operator is
\beq
V(k ; \hat{g_{ab}}) = \kappa_{dil} \int d^2 z \sqrt{\hat{g}}
\{ \hat{ g^{z \bar{z}} } \partial_z  X^M \partial_{\bar{z}} X^N 
\epsilon^{dil}_{MN}+ \frac{\alpha'}{4} \epsilon^{dil \;M }_{\;\;\;\;\;\;\;\;M}
 R^{(2)}(\hat{g}) \} \;\; .
\eeq
The polarization tensor of the dilaton is
$
\epsilon^{dil}_{MN} = (\eta_{MN} - k_M \bar{k}_N - k_N \bar{k}_M)
/ \sqrt{D-2}
$
where $k \cdot k = \bar{k} \cdot \bar{k} =0$ and $k \cdot \bar{k}=1$.
The torus n-point amplitude is
\beqn
& & \tilde{A}^{(n)}_{torus} (k_1, \cdots , k_n)
 \nonumber\\
 &=&
	\frac{1}{Vol(CKV)}
	(2 \pi)^D  \delta^{(D)} \left( \sum_{I=1}^{n} k_I \right)
	\int d^2 \tau \frac{1}{(\tau_2)^{14}}
		 \prod_{m=1}|1-e^{2 \pi i m \tau}|^{-48} e^{4 \pi \tau_2}
 \nonumber\\
 & &
	(\kappa_{dil})^n \prod_{I=1}^{n}
	 \int \sqrt{(\tau_2)^2} d^2 z_I \; \epsilon^{dil \;M_I N_I}
	 \frac{\partial}{\partial \bar{\eta}_{(I) N_I}}
	 \frac{\partial}{\partial \eta_{(I) M_I}}
 \nonumber\\
 & &
\exp \left[ \pi \alpha' \sum_{IJ}^n 
	(k_I - i \eta_{(I)} \partial_{z_I}
 - i \bar{\eta}_{(I)} \partial_{\bar{z}_I})
	\cdot
	(k_J - i \eta_{(J)} \partial_{z_J}
 - i \bar{\eta}_{(J)} \partial_{\bar{z}_J})
 	G_{torus}(z_I,z_J) 
\right.
 \nonumber\\
 & & 
	\left. \left.
	+ \frac{\alpha'}{4} \sum_{I}^n \eta_{(I)} \cdot \bar{\eta}_{(I)}
	R^{(2)}(z_I) \right] \right|_{\eta = \bar{\eta}=0}
\eeqn
where the Green's function on torus is given by
\beq
 G_{torus}(z_I,z_J) = \frac{1}{4 \pi} \log 
	\left| 
	\frac{{\cal \theta}_1(z_I-z_J|\tau)}{{\cal \theta}_1'(0|\tau)}
	\right|^2
+ \frac{i}{4} \left\{ 
		\frac{(z_I-z_J -\bar{z}_I+\bar{z}_J)^2}{\tau-\bar{\tau}}
	      \right\} 
\;\;.
\eeq
From the above expression we find
\beqn
\tilde{A}^{(1)}_{torus} (k)
&=&
- (2 \pi)^{D}  \delta^{(D)} \left(  k \right)
	\frac{\alpha'}{2} \pi \kappa_{dil}
	 \epsilon^{dil \;M }_{\;\;\;\;\;\;\;\;M}
	\int d^2 \tau \frac{1}{(\tau_2)^{14}}
		 \prod_{m=1}|1-e^{2 \pi i m \tau}|^{-48} e^{4 \pi \tau_2}
\;\;,
\nonumber
\eeqn
  Fourier transform  of which is   constant:
\beqn
A^{(1)}_{torus} (x)
&=&
	- \frac{\alpha'}{2} \pi \sqrt{D-2} \kappa_{dil}
	 \Lambda \;\;.
\eeqn
  Here
\beq
\Lambda \equiv
	\int d^2 \tau \frac{1}{(\tau_2)^{14}}
		 \prod_{m=1}|1-e^{2 \pi i m \tau}|^{-48} e^{4 \pi \tau_2}
\eeq
is the vacuum amplitude  of torus \cite{Pol1}.
The solution to  eq.~(\ref{eqn:fs_tadpole_cancel}) is
\beq
 j_{cl}(x) = - \frac{\alpha'}{2} \pi (\sqrt{D-2}) 
	   \Lambda \; x^{M}x_{M}
\eeq
  to the order we are considering.
 The loop-corrected  consistent background metric is
\beq
G_{MN}(x^{M}) = \eta_{MN} 
	\left( 1 - (\kappa_{dil})^2 \frac{\alpha'}{2} \pi 
	 (\sqrt{D-2})  \Lambda \; x^{M}x_{M}
	\right).
\eeq
   This is in agreement  with \cite{FS}.
  Although we will not discuss here,  one can proceed to the cancellation
  of infinities in  the $n$-point amplitude on this background
 which is given by
\beq
 F.T. \;   \left.  
  \frac{\delta^n}{\delta j(x_1) \cdots \delta j(x_n)} 
	\left( {\cal S}_{sphere} [j] + {\cal S}_{torus} [j] \right)
 \right|_{j = \kappa_{dil} j_{cl}(x)} \;\;.
\eeq

\end{document}